\documentclass[reprint,amsmath,amssymb,aps,pra,longbibliography]{revtex4-1}
\usepackage{amsmath}
\usepackage{braket}
\usepackage{graphicx}
\usepackage[shortlabels]{enumitem}
\usepackage{substr}
\usepackage{verbatim}
\usepackage{import}
\usepackage{siunitx}
\usepackage{mathtools}
\usepackage{hyperref}
\usepackage{lipsum}
\usepackage{float}
\usepackage{xfrac}
\hypersetup{colorlinks=true, citecolor=blue, urlcolor=blue, linkcolor=blue}
\allowdisplaybreaks  

\begin{document}
\title{Comment on ``Matter-wave interferometry with helium atoms in low-$l$ Rydberg states''}
\author{D.~Z.~Chan}
\thanks{Present address: \emph{Department of Physics and John Adams Institute for Accelerator Science, University of Oxford, Denys Wilkinson Building, Keble Road, Oxford OX1 3RH, United Kingdom}, \href{mailto:darren.chan@physics.ox.ac.uk}{darren.chan@physics.ox.ac.uk} and \href{mailto:dzchan@uwaterloo.ca}{dzchan@uwaterloo.ca}}
\author{J.~D.~D.~Martin}
\affiliation{Department of Physics and Astronomy, University of Waterloo, Waterloo N2L 3G1 Canada}
\date{\today}
\begin{abstract}
Tommey and Hogan [\href{https://journals.aps.org/pra/abstract/10.1103/PhysRevA.104.033305}{Phys. Rev. A \textbf{104}, 033305 (2021)}] have reported a matter-wave interference experiment using Rydberg atoms traveling through inhomogeneous electric fields at $\approx \SI{2000}{m/s}$.  Using a simplified model containing the essential physics of their experiment, we show that the phase difference measured by their observed interference fringes does not depend --- in any significant way --- on the acceleration of the Rydberg atoms, but instead simply on the uniform motion of the atoms through the inhomogeneous electric field.
\end{abstract}
\maketitle

Tommey and Hogan (TH) \cite{shortdoi:gmsf2r} have recently reported experimental observations of matter-wave interferometry using a superposition of two Rydberg states of helium: $1s56s^3S_1$ and $1s57s^3S_1$ (``$g$'' and ``$e$'' hereafter). They emphasize the importance of inhomogeneous electric fields in their experiment, which in principle cause different accelerations of the two internal states.  These ``Stark accelerations'' are due to the forces that electric dipoles experience within inhomogeneous electric fields.  Since the two states have different polarizabilities, they will have different dipole moments in the same electric field, and thus they experience different forces and corresponding accelerations.

The purpose of this Comment is to clarify an aspect of TH's observations that might otherwise lead to misinterpretation.  Specifically, we derive closed-form expressions for the phases measured by their interferograms.  By numerical evaluation of these expressions, we show that their observed phases are primarily due to Stark \emph{shifts} instead of \emph{accelerations}, clarify the role of their large beam velocity, and briefly discuss whether their observations should be considered to be matter-wave interferometry.

\begin{figure*}
\begin{center}
\includegraphics{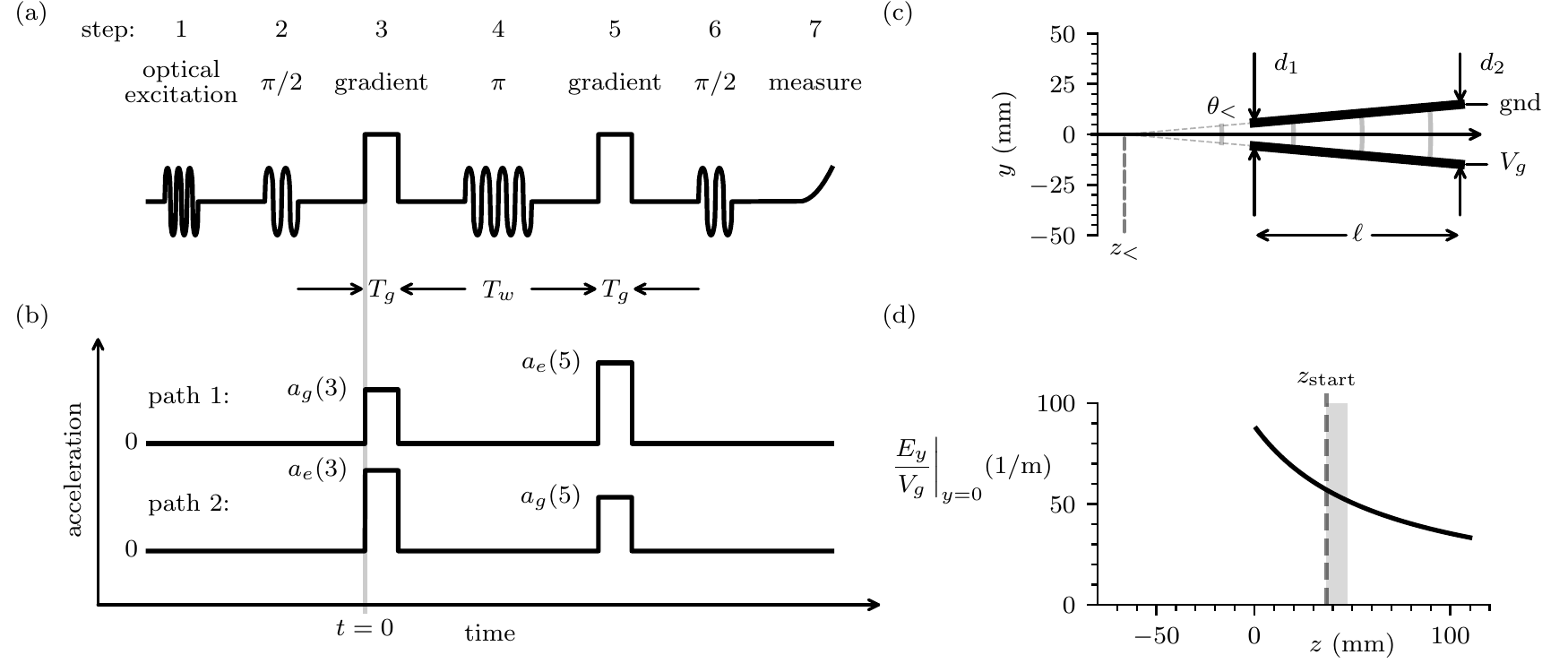}
\end{center}
\caption{(a) \label{fg:electrodes_timing} Steps in the interferometry experiment, where $T_g = \SI{160}{ns}$, and $T_w = \SI{5000}{ns}$.
(b) \label{fg:acceleration_comparison} Accelerations corresponding to the two paths through the interferometer: $a_g(3)$ and $a_e(3)$ are the accelerations during the gradient pulse in Step \ref{it:gradient}, and $a_g(5)$ and $a_e(5)$ are the accelerations during the gradient pulse in Step \ref{it:gradient again}.
(c) \label{fg:electrode_geometry} The electrode geometry used to generate electric field gradients in TH, with $\theta_< \approx \SI{0.17}{rad}$ and $z_< \approx \SI{-66.4}{mm}$, as determined by $d_1 = \SI{11.5}{mm}$, $d_2 = \SI{29.7}{mm} $, and $\ell = \SI{105}{mm}$.
(d) \label{fg:Ey_variation} The variation in $E_y$ along the $z$-axis ($y=0$), normalized by $V_g$ (note that $E_z=0$ along the $z$-axis).  The shaded area indicates the range of $z$ distances that we assume the atoms to travel over from the start of first gradient pulse (Step \ref{it:gradient}), $z_\text{start} = \SI{36.7 \pm 1}{mm}$, until the end of the second gradient pulse (Step \ref{it:gradient again}).
}
\end{figure*}

The experiments performed by TH consist of repeating the following steps (see Fig.~\ref{fg:electrodes_timing}(a)):
\begin{enumerate}
\itemsep0em
    \item \label{it:excitation} prepare a sample of Rydberg atoms, each in state $g$;
    \item \label{it:pi/2} form a superposition of states $g$ and $e$ using a resonant $\pi/2$ microwave pulse;
    \item  \label{it:gradient} expose the atoms to a ``gradient'' electric field
    that causes the two internal state center-of-mass (COM) wavefunctions formed in Step \ref{it:pi/2} to split;
    \item \label{it:pi} apply a $\pi$ pulse during a waiting period to ``swap'' the internal state of each COM wave function;
    \item \label{it:gradient again} apply a second gradient pulse with the same properties as in Step \ref{it:gradient};
    \item \label{it:pi/2 again} apply a resonant microwave $\pi/2$ pulse identical to Step \ref{it:pi/2}; then
    \item measure the number of Rydberg atoms in each of the two states by state-selective field ionization \cite{shortdoi:gr3mdk}.
\end{enumerate}
The microwave source should be coherent across Steps \ref{it:pi/2}, \ref{it:pi}, and \ref{it:pi/2 again}.

If we assume that the $\pi$ pulse in Step \ref{it:pi} perfectly swaps $e$ and $g$, then there are two well-defined ``paths'' through this interferometer from the first to the last $\pi/2$ pulses.   At the end of the sequence, one detects a mixture of $e$ and $g$ states that depends on the interference between these two paths. Interferograms are collected by measuring the $e$-state population $P_e = [1 - C \cos (\Delta \phi)]/2$, where $C$ is the contrast \footnote{Our work here does not address the loss of contrast with increasing phase, but could be extended to address it using the same approach as TH; i.e., averaging over positions and longitudinal velocity distributions.}, as a function of an electrode potential that generates the field in Steps \ref{it:gradient} and \ref{it:gradient again}. Figure 3(a) of TH is an interferogram collected in this manner, clearly exhibiting interference fringes; but as we will show, these fringes do not primarily depend on Stark acceleration.

Let us review how TH compute $\Delta \phi$ for the comparisons that they make with their observations.  An ``action phase'' for each path ($i=1$ or $2$) is decomposed into a dynamic and Stark phase, $\phi_{\text{action},i} = \phi_{\text{dynamic},i} - \phi_{\text{Stark},i}$, with
\begin{equation} \label{eq:dynamic}
\phi_{\text{dynamic},i} = \frac{m}{2\hbar} \int_{0}^{t} v_i^2 \: dt',
\end{equation}
where $m$ is the mass of the helium atom, and
\begin{equation} \label{eq:stark}
\phi_{\text{Stark},i} = \frac{1}{\hbar} \int_{0}^{t} \Delta W_{S_i(t')}(t') \: dt',
\end{equation}
where $\Delta W_{S_i(t')} (t')$ is the Stark energy shift of an atom in state $S$ ($e$ or $g$).  An explicit time dependence is given for $S_i$ because the internal state changes at the $\pi$ pulse.  The ``separation phase'' is given by:
\begin{equation} \label{eq:sep_phase}
\Delta \phi_{\text{separation}} = \frac{1}{\hbar}
\left(
\frac{p_1+p_2}{2}
\right)
\left(z_1-z_2 \right),
\end{equation}
where $p_i$ and $z_i$ are the final computed momenta and positions for each path. The variables $p_i$, $z_i$ and the integrals in Eqs.~\ref{eq:dynamic} and \ref{eq:stark} are evaluated using the classical trajectories for the COM motion.

The final phase difference is written as:
\begin{equation} \label{eq:total_phase}
\Delta \phi = \Delta \phi_{\text{dynamic}} + \Delta \phi_{\text{separation}} - \Delta \phi_{\text{Stark}},
\end{equation}
where $\Delta \phi_{\text{dynamic}} = \phi_{\text{dynamic},2} - \phi_{\text{dynamic},1}$ and similarly for $\Delta \phi_{\text{Stark}}$.  Note that the path indices in the subtraction are in the reverse order from those in Eq.~\ref{eq:sep_phase} \cite{shortdoi:gr4dv4}.

In the absence of any acceleration in Steps \ref{it:gradient} and \ref{it:gradient again}, it is straightforward to show that $\Delta \phi_{\text{dynamic}} + \Delta \phi_{\text{separation}} = 0$ \cite{ourgithubrepo}.  On the other hand, if the electric field differs in Steps \ref{it:gradient} and \ref{it:gradient again}, then in general $\Delta \phi_{\text{Stark}} \ne 0$, even with no acceleration during these steps. For this reason, it is helpful to define a ``matter-wave'' phase as
$\Delta \phi_{\text{matter}} \coloneqq \Delta \phi_{\text{dynamic}} + \Delta \phi_{\text{separation}}$, which corresponds to Stark {\em acceleration}, whereas $\Delta \phi_{\text{Stark}}$ corresponds to Stark {\em shifts}.

For instance, if Steps \ref{it:gradient} and \ref{it:gradient again} are of duration $T_g$ with a waiting period of $T_w$ in-between (see Fig.~\ref{fg:acceleration_comparison}(b)), then Eq.'s \ref{eq:dynamic} and \ref{eq:sep_phase} may be used to determine \cite{ourgithubrepo}:
\begin{equation} \label{eq:matter_phase}
\Delta \phi_{\text{matter}} =
  \frac{m}{2 \hbar} \: (a_{g}^{2} - a_{e}^{2}) \: T_{g}^{2} (T_{g} + T_{w}),
\end{equation}
where we have assumed 1) instantaneous $\pi/2$ and $\pi$ pulses, and that 2) the Stark acceleration of an atom in the $g$ state is the same in Steps \ref{it:gradient} and \ref{it:gradient again} ($a_g(3) = a_g(5) \eqqcolon a_g$) and likewise for the $e$ state.  Expressions similar to Eq.~\ref{eq:matter_phase} are given in Ref.'s \cite{shortdoi:gpb7cs,shortdoi:gn7jtg,shortdoi:gptnmb} for slightly different interferometry sequences.

Note that $\Delta \phi_{\text{matter}}$ as given by Eq.~\ref{eq:matter_phase} is invariant under Galilean transformations between inertial reference frames, whereas the individual contributions, $\Delta \phi_{\text{dynamic}}$ and $\Delta \phi_{\text{separation}}$, are not invariant, suggesting caution in considering them individually.

We now estimate the value of $\Delta \phi_{\text{matter}}$  using Eq.~\ref{eq:matter_phase}, by determining $a_g$ and $a_e$. Accelerations are generated on atoms in state $S$ via $\mathbf{a}_S = -[\boldsymbol{\nabla} (\Delta W_S) ] / m$ where $m$ is the mass of a helium atom.  For simplicity, we will assume that the Stark shifts are quadratic in the electric field magnitude $E$, so that:
\begin{equation} \label{eq:quad_stark}
\Delta W_{S} = -\frac{1}{2} \alpha_S E^2
\end{equation}
where $\alpha_S$ is the polarizability of either $e$ or $g$ state, as given by TH.

To generate an electric field gradient TH use two electrodes in a ``wedge''-type configuration (see Fig.~\ref{fg:electrode_geometry}(c)).
As the electrodes are much wider in the $\mathbf{\hat{x}}$ direction than their separation in the $\mathbf{\hat{y}}$ direction, we may neglect any fields or variations in the $\mathbf{\hat{x}}$ direction. If for simplicity, we consider only atoms moving along the $z$-axis with $y=0$, then in this geometry we find
$\boldsymbol{\nabla} (E^2)|_{y=0} = 2 (E_y \partial_z E_y)|_{y=0} \: \mathbf{\hat{z}}$.

To determine $(E_y \partial_z E_y)|_{y=0}$, we adopt a similar model to TH, writing \footnote{To obtain TH's model, replace $\theta_<$ by $2\tan (\theta_< / 2)$ in Eq.~\ref{eq:on_axis_model}, with a negligible difference for small $\theta_<$ such as $\theta_< \approx \SI{0.17}{rad}$.
}:
\begin{equation} \label{eq:on_axis_model}
   E_y |_{y=0} = \frac{V_g}{\theta_< (z-z_<)},
\end{equation}
where $V_g$ is the potential applied to the bottom electrode, the top electrode is grounded, and $\theta_<$ and $z_<$ are defined in Fig.~\ref{fg:electrode_geometry}(c).  This approximate model may be justified by considering an infinite extension of the two electrodes in the $+\mathbf{\hat{z}}$, $-\mathbf{\hat{z}}$, $+\mathbf{\hat{x}}$, and $-\mathbf{\hat{x}}$ directions to form an ``infinite wedge'', resulting in a two-dimensional problem in the coordinates $y$ and $z$.  The corresponding solution of Laplace's equation for the potential in this geometry is a standard result of complex variable theory
\cite{isbn:9780139078743} and gives Eq.~\ref{eq:on_axis_model} for $E_y |_{y=0}$ along the $z$-axis.

We now consider the time-dependence of the bottom electrode voltage that determines the gradients in Steps \ref{it:gradient} and \ref{it:gradient again}.
Equation \ref{eq:matter_phase} assumes constant accelerations during the gradient pulses, whereas TH apply a time-dependent voltage to the bottom electrode (see Fig.~\ref{fg:electrode_geometry}(c)) by: 1) linearly ramping from zero to $V_{g}$ in \SI{130}{ns}, 2) holding constant $V_{g}$ for \SI{72}{ns}, and finally 3) linearly ramping from $V_{g}$ to zero in \SI{130}{ns}.
To simplify, we instead consider voltage pulses of the same value of $V_{g}$, and thus the same peak accelerations, but with negligible rise and fall times (see Fig.~\ref{fg:electrodes_timing}(a)).  We choose a duration of $T_{g} = \SI{160}{ns}$ to keep $\int (E_y(t))^2 \: dt$ roughly the same over each pulse as for TH allowing us to compare Stark phases with TH.

With the field model of Eq.~\ref{eq:on_axis_model}, we find the state-dependent accelerations:
\begin{equation} \label{eq:accel}
\mathbf{a}_S = - V_g^2 \: \frac{\alpha_S}{m}
\frac{1}{(z-z_<)} \left(
\frac{E_y|_{y=0}}{V_g}
\right)^2
\: \mathbf{\hat{z}}
\end{equation}
with $E_y|_{y=0}/V_g$ depending solely on $z$ and the geometry of the electrodes, as illustrated in Fig.~\ref{fg:Ey_variation}(d).

Not all of the atoms have the same trajectories due to their different initial velocities, and the finite duration and spatial extent of the optical excitation in Step \ref{it:excitation}. We focus on the trajectory of an atom with the average initial beam velocity of $v_{\text{beam}} = \SI{2000}{m/s}$. Since $E_y|_{y=0}/V_g$ varies along the $z$-axis, we need a model to determine where this atom is along the $z$-axis at any point in the experimental sequence. Since the accelerations are weak (\emph{vide infra}), we take $z = z_{\text{start}} + v_{\text{beam}} t$, estimating that $z_\text{start} \approx \SI{36.7 \pm 1}{mm}$ at the start of the first gradient pulse, which we define as $t=0$. (We assume that the location of optical excitation is the same as given in Ref.~\cite{shortdoi:gg2bbk}.)

At $t = 0$, we find that $\mathbf{a}_{56s} / V_g^2 \approx \SI{-660}{m/(s^2 V^2)}\: \mathbf{\hat{z}}$, and thus over the approximately $\SI{160}{ns}$ duration of a $V_g = \SI{1}{V}$ gradient pulse, the atoms would only change velocity by $\Delta v_z \approx \SI{-e-4}{m/s}$, which is small in magnitude compared to $v_{\text{beam}} = \SI{2000}{m/s}$.

We now have the information required to estimate $\Delta \phi_{\text{matter}}$ as given by Eq.~\ref{eq:matter_phase}. If we model $a_g$ as the average of $a_g(3)$ and $a_g(5)$ (Eq.~\ref{eq:accel} evaluated at the midpoints of each of the gradient pulses), and similarly for $a_e$, we find that $(a_g^2-a_e^2) / V_g^4 \approx -\SI{1.0e5}{m^2/(s^4 V^4)}$. From Eq.~\ref{eq:matter_phase}, we may estimate $\Delta \phi_{\text{matter}}/(2\pi) \approx -\SI{3e-5}{}$ for $V_g \approx \SI{4.5}{V}$, which is the largest $V_g$ in the example interferogram of Fig.~3(a) of TH.  In contrast, by counting fringes in this interferogram, we estimate $\Delta \phi /(2\pi) \approx 11$ for $V_g \approx \SI{4.5}{V}$.

In using Eq.~\ref{eq:matter_phase}, we have assumed that $a_g(3)=a_g(5)$ and $a_e(3) = a_e(5)$ (see Fig.~\ref{fg:acceleration_comparison}(b)). But because of the position dependence of $(E_y \partial_z E_y)|_{y=0}$ and the motion of the atoms, these accelerations cannot be equal during both gradient pulses (or even within a gradient pulse). However, introducing differing accelerations for the two gradient pulses does not significantly change the estimate for $\Delta \phi_{\text{matter}}$. Defining $f$ so that $a_g(5) = (1+f) a_g(3)$ and $a_e(5) = (1+f) a_e(3)$, we find that the $T_g+T_w$ factor in the RHS of Eq.~\ref{eq:matter_phase} should be replaced by
$\left[
T_{g} \left(1 + \frac{2}{3} f - \frac{1}{6} f^2 \right)
+ T_{w} (1 + f)
\right]$,
which reduces to $T_g+T_w$ when $f=0$ (as required).
Using the field and acceleration models (Eqs.~\ref{eq:on_axis_model} and \ref{eq:accel}), together with assumed motion of the atoms, we estimate $f \approx -0.25$, not significantly changing our small estimate for $|\Delta \phi_{\text{matter}}|$ \footnote{Variations of acceleration \emph{within} each gradient pulse will be small in comparison to \emph{between} pulses, due to the small gradient pulse times ($T_g$) compared to their separation ($T_w$).}.

A small fractional change in $|\Delta \phi_{\text{matter}}|$ is also obtained if we consider the two gradient pulses to have slightly different durations \cite{ourgithubrepo}.   Likewise, it is not expected that our idealization of TH's gradient pulses changes $|\Delta \phi_{\text{matter}}|$ significantly.

In principle, TH's interferograms measure $\Delta \phi = \Delta \phi_{\text{matter}} + \Delta \phi_{\text{Stark}}$. In practice, given our estimate of a negligible $|\Delta \phi_{\text{matter}}|$ and the observed signal to noise, TH's interferograms measure $\Delta \phi_{\text{Stark}}$, which we now estimate.   For a non-zero $\Delta \phi_{\text{Stark}}$, it is sufficient that the electric field experienced during Steps \ref{it:gradient} and \ref{it:gradient again} differ --- a field gradient during Steps \ref{it:gradient} and \ref{it:gradient again} is \emph{not} required for a non-zero Stark phase. Thus, we use the field magnitudes computed at the midpoint of the two gradient pulses, calling them $E(3)$ and $E(5)$.  Using Eqs.~\ref{eq:stark} and \ref{eq:quad_stark}, we find:
\begin{equation} \label{eq:stark_phase}
\Delta \phi_{\text{Stark}} =
 - V_g^2 \frac{T_g}{2 \hbar} \left( \alpha_e - \alpha_g \right)
\left[
\left(
\frac{E(3)}{V_g}
\right)^2
-
\left(
\frac{E(5)}{V_g}
\right)^2
\right]
\end{equation}
where $[(E(3)/V_g)^2-(E(5)/V_g)^2] \approx \SI{5.4e2}{m^{-2}}$, as evaluated using Eq.~\ref{eq:on_axis_model} and our model for the motion of the atoms during the pulse sequence.

Considering our idealizations, Eq.~\ref{eq:stark_phase} reproduces TH's observed $\Delta \phi$ reasonably well (in the quadratic Stark shift regime). For example, from the interferogram in Fig.~3(a) of TH, we estimate that $\Delta V_g \approx \SI{0.65\pm0.02}{V}$ is required to go from the central null to the first interference maxima.  By setting the left hand side of Eq.~\ref{eq:stark_phase} to $\pi$ and rearranging, we find $\Delta V_g \approx \SI{0.62\pm0.01}{V}$ (where the error estimate is based on a \SI{1}{mm} uncertainty in $z_{\text{start}}$).

The preceding analysis shows that the fringes in TH's interferograms depend primarily on inertial motion from one region of the wedge to another --- Stark accelerations play an insignificant role and are not necessary to observe the interference fringes
\footnote{The effect of the two paths sampling the field in different locations due to their different accelerations is small since --- as estimated earlier --- the $\Delta v$'s arising from Stark acceleration ($\approx -\SI{e-4}{m/s}$) are much smaller than $v_{\text{beam}} \approx \SI{2000}{m/s}$.)}.

To reinforce the preceding point, imagine an experiment with atoms at rest in the lab, located between two parallel electrodes that generate \emph{homogeneous} electric fields.
The experimental sequence would be the same as TH (Fig.~\ref{fg:acceleration_comparison}(a)) with the electric fields generated in Steps \ref{it:gradient} and \ref{it:gradient again} matching the values of $E(3)$ and $E(5)$ in TH's experiment for a given $V_g$ (where the fields sampled by the atoms differ because of their motion through the inhomogeneous field).  The phase measured in this hypothetical experiment would be given by Eq.~\ref{eq:stark_phase}, and thus would be the same as TH's observed phase (for a given $V_g$), since $|\Delta \phi_{\text{matter}}|$ is so small in comparison to $|\Delta \phi_{\text{Stark}}|$.  In this hypothetical experiment, Stark \emph{acceleration} is not required to observe interference.  Similarly, Stark acceleration does not play a significant role in determining the phases measured by TH's interferograms.

A negligible $\Delta \phi_{\text{matter}}$ is consistent with Figure 3(b) of TH, which shows --- at least approximately --- that $\Delta \phi_{\text{separation}}$ and $\Delta \phi_{\text{dynamic}}$ tend to cancel.  TH comment on this cancellation, suggesting that it is related to the large speed that their atoms have in the lab frame and that lower speeds may be beneficial.  However, the $\Delta \phi_{\text{matter}}$ expression given by Eq.~\ref{eq:matter_phase} is \emph{independent} of any initial motion that the atoms in the $\mathbf{\hat{z}}$ direction.  (A non-zero initial $v_z$ is assumed in the derivation but cancels out of the final expression \cite{ourgithubrepo}.)  This independence suggests that --- all other things being equal --- it would be difficult to observe $\Delta \phi_{\text{matter}} \ne 0$ in TH's experiment by simply reducing the beam velocity.

Finally, we consider the use of the adjective ``matter-wave'' to describe TH's experiments. TH observe interference fringes with increasing phase until the separation of the wavepackets due to Stark acceleration is roughly $|\Delta z| \approx \SI{0.75}{nm}$.  They compare this separation with the mean de Broglie wavelength of the atoms in the laboratory frame $\lambda_{\text{dB}}$, finding that $|\Delta z| \approx 15 \lambda_{\text{dB}}$. However, by a Galilean transformation into an inertial reference frame in which the atoms move more slowly, $\lambda_{\text{dB}}$ can be arbitrarily lengthened, exceeding $|\Delta z|$, which remains unchanged.  As such, $|\Delta z| \gtrsim \lambda_{\text{dB}}$ is not a criteria for matter-wave interferometry.

Alternately, one may compare the separation $|\Delta z|$ to the \emph{thermal} de Broglie wavelength $\lambda_{\text{th}}$ in the (mean) rest frame of the atoms.  For a classical gas, $\lambda_{\text{th}}$ is the coherence length for interference between spatially separated parts of a wavefunction \cite{shortdoi:b4gn4k}.  For TH's experiment $\lambda_{\text{th}} \approx \SI{0.8}{nm}$ \footnote{TH's simulations sample from $z$-axis velocity distribution with a standard deviation of $\sigma = \SI{50}{m/s}$, as determined by time-of-flight measurements.  This distribution corresponds to a temperature of $T = m \sigma^2 / k_B \approx \SI{1.2}{K}$, so that $\lambda_{\text{th}} = \sqrt{2\pi\hbar^2/(m k_B T)} \approx \SI{0.8}{nm}$ }, which is close to $|\Delta z| \approx \SI{0.75}{nm}$, suggesting that although Stark acceleration is not required for observation of the fringes, it may be responsible for their disappearance with increasing phase.

Thus, whether or not one considers TH's experiment to be a demonstration of matter-wave interferometry might depend on how one thinks Stark acceleration should impact TH's interferograms.  It is not unreasonable to expect that a \emph{matter-wave} interferometer should produce measurements that depend on the existence of two \emph{distinct} paths through spacetime (in this case created by Stark acceleration).  In this Comment, we have demonstrated that the \emph{phases} measured in TH's interferograms do not depend on there being two distinct paths through spacetime. (That the paths slightly differ hardly matters
\footnote{A similar situation is as follows:  imagine a conventional Ramsey interferometer, with no accelerations as normal. Then, suppose field gradients are introduced which cause the two internal states to accelerate slightly differently. Unless these accelerations are noticeable in the phases measured by the interferograms, one would not refer to this Ramsey interferometer as a matter-wave interferometer.}.)
On the other hand, since $|\Delta z| \approx \lambda_{\text{th}}$ the reduction in the \emph{contrast} of TH's interferograms with increasing phase is possibly due to the existence of two paths.

In summary: by deriving analytical expressions for the phases measured by TH's interferograms, we have shown that Stark accelerations (Eq.~\ref{eq:matter_phase}) make a numerically negligible contribution to their measured phases, whereas Stark shifts (Eq.~\ref{eq:stark_phase}) dominate.  Similar considerations may apply to the Rydberg atom interferometry experiment reported in Ref.~\cite{shortdoi:gg2bbk}.

We thank H.~J.~Kim and A.~Kumarakrishnan for helpful discussions.  This work was supported by NSERC (Canada).

\bibliography{002_references+references,003_references+extras}

\end{document}